\documentclass[times, 11pt,latex8]{article}

\usepackage{amsmath}
\usepackage{amssymb}
\usepackage{amsfonts}
\usepackage{amstext}
\usepackage{latexsym}
\usepackage{epsfig}
\usepackage{graphicx}

\newtheorem{propn}{Proposition}{}{}
\newtheorem{thm}{Theorem}{}{}
\newtheorem{lem}{Lemma}{}{}
\newtheorem{emptyT}{}{}{}
\newtheorem{cor}{Corollary}{}{}
\newtheorem{Def}{Definition}{}{}
{}{}
{}{}
\newenvironment{proof}{{\bf Proof:}}{~$\Box$ \\}
\newcommand{\commentout}[1]{}

\newcommand{\Complex}{\mathbb{ C}}

\newcommand{\norm}[1]{|\!| #1 |\!|}

\newcommand{\ket}[1]{|#1\rangle}

\newcommand{\bra}[1]{\langle #1|}
\newcommand{\inp}[2]{\langle #1|#2\rangle}
\newcommand{\inpr}[3]{\langle #1|#2|#3\rangle}

\newcommand{\udist}{{\em u}-distance}

\newcommand{\conj}[1]{\overline{#1}}
\newcommand{\hconj}[1]{{#1}^{\dagger}}

\newcommand{\tr}{{\tt Tr}}
\newcommand {\pj}[1]{\ket{#1}\bra{#1}}

\newcommand{\bcolvec}{\left(\begin{array}{c}}
\newcommand{\ecolvec}{\end{array}\right)}
\newcommand{\bmat}[1]{\left(\begin{array}{#1}}
\newcommand{\emat}{\end{array}\right)}
\newcommand{\imp}{\Rightarrow}

\newcommand{\tensor}{\otimes}

\newcommand{\be}{\begin{enumerate}}
\newcommand{\ee}{\end{enumerate}}
\newcommand{\beq}{\begin{equation}}
\newcommand{\eeq}{\end{equation}}
\newcommand{\beqx}{\begin{displaymath}}
\newcommand{\eeqx}{\end{displaymath}}
\newcommand{\beqa}{\begin{eqnarray}}
\newcommand{\eeqa}{\end{eqnarray}}
\newcommand{\beqax}{\begin{eqnarray*}}
\newcommand{\eeqax}{\end{eqnarray*}}

\newcommand{\field}[1]{{\mathbb #1}}
\newcommand{\cali}[1]{{\mathcal #1}}

\begin{document}
\title{Projective invariant measures and approximation of quantum circuits}

\author{
M.~K. ~Patra \\
CQCT, Macquarie University, NSW 2019,  Australia \\
manas@ics.mq.edu.au
}
%%\institute{}
\date{}
\maketitle

\begin{abstract}
In this paper we introduce a projective invarinat measure on the special unitary group. It is directly related to transition probabilities. It has some interesting connection with convex geometry. Applications to approximation of quantum circuits and entanglement are given. 
\end{abstract}

\section{Introduction}
The concept of approximation of one operator by another is important
in many branches of physics and mathematics. For quantum computing and
information in particular, the approximation of unitary or more
general operators is crucial for implementation of quantum algorithms.
In any approximation theory the notion of distance is essential for a
quantitative estimate of the accuracy of approximation. In the
approximation of linear operators on a Banach( or Hilbert) space, the
usual distance function or metric is induced by the norm. This
approach is particularly useful if we restrict to affine subspaces of
the space of operators. This is because the norm on the ambient space
induces a norm on the space of operators and the metric is defined in
terms of the latter. However, if we restrict to some {\em subset} of
operators which may not constitute a subspace, the norm induced metric
may not seem very natural. For two operators $A$ and $B$ the
difference $A-B$ whose length defines the distance between $A$ and $B$
may take us outside the subset. But the concept of a metric does not
depend upon algebraic operations. In particular, if the relevant subset
is a group we are often interested in {\em invariant} metrics. That is
metrics that remain invariant under left (right) translations by the
group operations. Of course, invariant metrics are known to exist for any compact group.

We may also view the problem of approximation of an operator from another perspective. Informally, one could say that a sequence of
unitary operators $U_n$ converges to some unitary operator $U$ if, for
any given state $\ket{\alpha}$ the expectation values of the sequence
$U_n$ converges to that of $U$. In fact we will adopt a weaker
criteria. Namely, that they converge in certain probability measures. This 
also turns out to be equivalent to convergence in the operator
norm. But first, a brief synopsis of the paper.

In Section II, I introduce a metric on the group of unitary
operators acting on a Hilbert space. The discussion will be
confined to an arbitrary but fixed Hilbert space, mostly finite
dimensional. Many algebraic and geometric properties of the metric
are proved. Some interesting connections with 2-polytopes (polygons)
are discussed. I then derive relations with other metrics. In
particular, equivalence with convergence in the operator norm is
shown. I also discuss approximation of quantum circuits.

In Section III, the definitions are extended to the case when the
ambient space has a tensor product structure. Some connection
between quantum state entanglement and the convex geometry of the
previous section is explored. I analyse bipartite entanglement from
a different perspective. I also discuss extensions to the difficult
case of multipartite entanglement.

I make some concluding remarks about some aspects
not covered in the paper which will be investigated subsequently.

\section{ An invariant metric on the special unitary group}
First, let us fix some notation. In the following, $\cali{H}$ will
denote a complex Hilbert space with a fixed inner product $<,>$. I also use
$\cali{H}_n$ to denote a space of dimension $n$. In the following the
dimension of all spaces under discussion will be assumed to be finite,
unless specified otherwise. Let $\cali{U}_n$ denote the group of
unitary opertaors in $B(\cali{H}_n)$, where the latter denotes the
algebra of linear operators on $\cali{H}_n$. The corresponding subset
of hermitian operators will be denoted by
$L(\cali{H}_n)$. The special unitary group $S\cali{U}_n\subset
\cali{U}_n$ is the subgroup of opertaors with determinant 1. I use the
standard notation \(\field{C} \text{ and } \field{R}\) for the field
of real and complex numbers with usual topology. In $\field{C}^n$ the
standard inner product is used. Thus, if \(\alpha = (x_1, \ldots, x_n)^T
\text { and } \beta = (y_1, \ldots, y_n)^T \in \field{C}^n\), where
$A^T$ denotes the transpose of the matrix $A$, then
\[ \inp{\alpha}{\beta} \equiv \sum_i \conj{x_i}y_i \]

 In this section $\cali{H}$ will denote a complex Hilbert space of
 dimension $n$. The Hilbert space norm induces a norm $A\rightarrow
 \norm{A}$, on the space of operators on $\cali{H}$, defined by,
\[ \norm{A} = \max_{\norm{\psi}=1} \norm {A\psi} =
 \max_{\norm{\psi},\norm{\phi}=1} \inpr{\phi}{A}{\psi}\]
 If $A$ is normal then \(\norm{A}= \max\{|\lambda| | \lambda \text{ an
 eigenvalue of $A$ } \). These and other properties of the norm
 $\norm{}$ may be found in \cite{Bhatia}. The norm induces a metric on
 the space of operators.

Let $\psi \in \cali{H}$ be a unit vector. For any pair
 of unitary operators $ U,V \in \cali{U}_n $ define
\beq \label{def:psi_dist}
D_{\psi} (U,V) = (1 - |\inpr{\psi}{U^{\dagger}V}{\psi}|^2 )^{1/2}
\eeq
Some of the obvious properties of $D_{\psi}$ are given in the
 following theorem.

\begin{thm} \label{thm:obvProp}
The function $D_{\psi}$ satisfies the following for any unitary
operators $U,V$. Let $W=\hconj{U}V$. .
\be
\item
\(0\leq D_{\psi}(U,V)  \leq 1 \). The first equality holds iff $\psi$
is an eigenvector of $W$.
\item
$D_{\psi}$ is left invariant with respect to group multiplication in
  $\cali{U}_n $. Thus, for any $X\in  \cali{U}_n $
\[D_{\psi}(XU,XV) = D_{\psi}(U,V) \]
Furher, \(D_{\psi}(UX, VX) = D_{X\psi} \). 
\item
$D_{\psi}$ is symmetric. That is, \(D_{\psi}(U,V)= D_{\psi}(V,U) \).
\item
$D_{\psi}$ satisfies the following
\beq \label{eq:basic_relation} 
\begin{split}
 \frac{1}{2}\norm{(U-e^{ix}V)\psi}^2 & \leq D_{\psi}^2(U,V) \leq  \norm{(U-V)\psi}^2 \\
 & \text{ for some real } x
 \end{split}
\eeq
\ee
\end{thm}
\begin{proof}
The first assertion follows from the Cauchy-Scwartz inequality and the
second and third from the definitions. We prove the last one. Note that 
\begin{equation*} 
\norm{(U-e^{ix}V)\psi}^2 = \inpr{\psi}({U-V)^{\dagger}(U-V)}{\psi} = 2(1-\text{Re}\inpr{\psi}{U^{\dagger} V} {\psi} 
\end{equation*}
where $\text{Re}(z)$ is the real part of the complex number $z$. 
There is a real number $x$ such that \(\inpr{\psi}{U^{\dagger} Ve^{ix}} {\psi} \) is positive. Then \(1-\inpr{\psi}{U^{\dagger} Ve^{ix}} {\psi} \leq 1- |\inpr{\psi}{U^{\dagger} Ve^{ix}}{\psi}|^2 = = D^2_{\psi}(U,V)\) and the first inequality in \ref{eq:basic_relation} follows. Moreover, 
\[
\begin{split}
D^2_{\psi}(U,V) & =  (1+ |\inpr{\psi}{U^{\dagger} V}{\psi}|)(1-|\inpr{\psi}{U^{\dagger} V}{\psi}|) \\
\leq 2(1-\text{Re}(\inpr{\psi}{U^{\dagger} V}{\psi}) 
\end{split}
\]
The last equality follows from the fact that $|\inpr{\psi}{U^{\dagger} V}{\psi}|\leq 1$. The theorem is proved. 
\end{proof}

As a simple illustration of the use of $D_{\psi}$ let us look at the
Grover search algorithm. A class of quantum algorithms which includes
the Grover algorithm may be reasonably described as an approximation
of a unitary operator by another with high probability. In the search
problem we are required to find a state labelled by a non-negative
integer $a$ say, from an unordered collection of such numbers. The
corresponding unknown operator we want to approximate may be taken to
be $U_a$, the operator that interchanges the basis states $\ket{0}$ and
$\ket{a}$ leaves the rest unchanged. Then the Grover algorithm
constructs a circuit, represented by a unitary operator $G_a$ , such
that $D_{\ket{0}} (U_a, G_a) < \epsilon$ for some error parameter
$\epsilon > 0$.

\begin{Def}
For any two unitary operators $U,V$ define
\beq \label{def:uDist}
D(U,V) = \max_{\norm{\psi}=1} D_{\psi}(U,V) = 1 -
\min{\norm{\psi}=1}|\inpr{\psi}{ \hconj{U}V} {\psi}|^2
\eeq
Call $D(U,V)$ the {\udist} between $U$ and $V$.
\end{Def}

I have used $max$( $min$) instead of $sup$( $inf$) in the above
definitions since the unit sphere \(S_{n-1}\equiv \{\psi\;| \;
  \norm{\psi}= 1\) is compact and the respective limits are
  attained. Now, define \(F_{\psi} (A) = \inpr{\psi}
    {A}{\psi} \) for any operator $A$. Then,
\[D(U, V) = 1- \min_{\norm{\psi}=1} |F_{\psi}( \hconj{U}V)|^2 \]
For an operator $A$, the set \( F(A) = \{F_{\psi}(A)|
\norm{\psi}=1\}\) is called the {\em field of values} or {\em
  numerical range} of $A$. It is a well-studied concept in linear
algebra \cite{HJ2}. We therefore have the first geometric
characteriztation of the {\udist}. Recall that for a metric space $(M,
\rho)$ with metric $\rho$ and for $x\in M \text{ and } K \subset S$,
the the distance between $x$ and $K$( also denoted by
$\rho$) is defined as
\[ \rho( x, K) = \inf_{y\in K} \rho(x,y) \]
If $K$ is compact then there exists $y\in K$ such that \(\rho( x, K)=
\rho( x, y) \).
\begin{emptyT}
$D(U,V)$ is the distance of the set $F(\hconj{U}V)$ from the
  origin. That is,
\[ D(U,V) = (1-\rho^2(0, F(\hconj{U}V)))^{1/2} \]
\end{emptyT}
Here the metric $\rho$ used is the standard Euclidean distance in
$\field{C}$.

Before listing the properties of $D(U,V)$ let us further investigate
its geometric meaning. First note that $D(U,V) = D(1, \hconj{U}V)
$. Hence it suffices to study the properties of $D(1,W)$ for a unitary
operator $W$. Since $W$ is unitary its eigenvalues lie on the unit
circle. If $z_i =e^{ic_i}$ is an eigenvalues of $W$,  then $0\leq c_i
< 2\pi$ and the angles are read counterclockwise on the unit
circle. Then writing an aribtrary vector in the basis of eigenvectors
of $W$ it is easy to see that the numerical range of $W$ is the {\em
  convex} set
\[F(W)= \{\sum_i |x_i|^2 e^{c_i}\; |\; \sum_i |x_i|^2 =1\}\]
That is, $F(W)$ is a convex 2-polytope or polygon whose vertices lie
on the unit circle. I derive below a simple expression for
$D(1,W)$. It depends upon a elementary geometric result that seems
obvious but the proof does not appear to be trivial. As I was unable
to find a published proof I give an elementary detailed one.
\begin{thm} \label{thm:eigenChar}
Let $z_i = e^{ic_i}, \; i= 1,\ldots n$ be the eigenvalues( possibly
with repititions) of a unitary operator $W$. Let $c_i$'s be ordered
such that $c_1\leq c_2\leq \cdots c_n$. Let $d = \max\{|c_i
-c_j|\}$. Then,
\beq
   D(1,W) = \begin{cases}
   |\sin (\frac{d}{2})| & \text{ if the $z_i$ lie
          inside a semicircle,}\\
           1  & \text{ otherwise.}
\end{cases}
\eeq	
\end{thm}
\begin{proof}
The theorem is intuitively obvious. Let $C$ be the the  arc connecting
the $z_i $'s.  If there are two arcs connecting all the eigenvalues
let $C$ be the smaller of the two arcs. If $C$ contains a semicircle
then the origin lies inside the polygon of $F(W)$. Hence, \(\min\;
\rho(0,F(W))= 0\). That is, $D(1,W) = 1$.  To make it more precise,
observe that \(D(1,W) = D(1, cW), \; |c|=1\). Hence, we may assume
that $c_1=0$. If $C$ includes a semicircle then it is either
lower or the upper semicircle. Suppose it is the upper
semicircle. Then there must be an eigenvalue on the upper semicirle
and another on the real axis or below it. Then the triangle joining
$c_1$ and these two eigenvalues contains the origin.

Next, suppose $C$ lies inside a semicircle, say the upper
semicircle. If it is any other semicircle then rotate $C$ by
multiplying with appropriate number $c, \; |c|=1$ so we get all the
eigenvalues in the upper semicircle. Then, $d= c_n$. Again it is
intuitively clear that the line joining the points 1 and $e^{ic_n}$
contains the point of the polygon that is closest to the origin. To
prove it directly we have to show that
\[ \min \{|p_1+\sum_{i=2}^n p_i e^{ic_i} |^2 : \; 0 \leq p_i\text{ and }
\sum_i p_i=1 \} = \cos^2(c_n/2) \]
I will follow essentially geometric intuition to prove this. First,
let $l$ be the line joining the points 1 and $z_n$. Then, it suffices
to prove that the line segment $l'$ joining the centre to an arbitray
point of  the polygon intersects $l$ at an {\em interior} piont of
$l'$. That
is, for any set \(\{p_1, \ldots, p_n| p_i\geq 0 \text{ and }\sum_ip_i
=1\}\) the equation
\beq \label{eq:intersect}
r\sum_i p_i e^{ic_i} = x e^{ic_1}+(1-x)e^{ic_n}, 
\eeq
has a unique solution with \(0\leq r \leq 1\text{ and } 0\leq x \leq 1\). As $c_1=0$ the above
equation is equivalent to the following pair of real equations.
\begin{flalign} \label{eq:pair} 
r(p_1+ \sum_{i=2}^n p_i\cos{c_i} ) & = x(1-\cos{c_n})+\cos{c_n} \\
r( \sum_{i=2}^n p_i \sin{c_i}) & = (1-x) \sin{c_n}
\end{flalign}
Hence
\begin{flalign*}
x&=\frac{r(p_1+ \sum_{i=2}^n p_i\cos{c_i} )- \cos{c_n}}{(1-\cos{c_n})}
\\
& = 1- r\frac{( \sum_{i=2}^n p_i \sin{c_i})} {\sin{c_n}} \\
\imp & r \frac{(p_1+ \sum_{i=2}^n p_i\cos{c_i} )} {(1-\cos{c_n})}
+\frac{( \sum_{i=2}^n p_i \sin{c_i})} {\sin{c_n}} =
  \frac{1}{1-\cos{c_n}}\\
\imp & r (p_1\sin{c_n}+ \sum_{i=2}^{n-1}p_i(\sin{(c_n
    -c_i)}+ \sin{c_i})+p_n\sin{c_n})= \sin{c_n} \\
\imp & r[p_1\sin{c_n} + 2\sin{(c_n/2)} (\sum_{n=2}^{n-1}p_i \cos{(c_n/2-
    c_i)} +p_n\cos{(c_n/2)}) = \sin{c_n} \\
\end{flalign*}
Since \(0=c_1\leq c_2 \leq \cdots \leq c_n <\pi \) we have \(
\cos{(c_n/2- c_i)} \geq \cos{(c_n/2)}\). Hence, on the left side of
the last equation the expression
\[p_1\sin{c_n} + 2\sin{(c_n/2)} (\sum_{n=2}^{n-1}\cos{(c_n/2-
    c_i)} +p_n\cos{(c_n/2)})\]
is greater than or equal to
\[ p_1\sin{c_n} + 2\sin{(c_n/2)} (\sum_{n=2}^{n}p_i\cos{(c_n/2)})=
\sin{c_n} \]
Consequently, $0\leq r \leq 1$. The fact that $r\geq 0$ follows from above since $\sin{x}$ is nonnegative in the upper semicircle. From the equation \ref{eq:pair} it follos that \(  0\leq x \leq 1 \) theorem is proved.
\end{proof}

\commentout{
First let
\[|p_1+\sum_{i=2}^n p_i e^{c_i} |^2 = (1+\sum_i p_i \cos{c_i})^2 +
(\sum_i \sin{c_i} )^2 = A^2+ B^2 \]
Since the points $z_i$ lie on the upper semicircle, $\sin{c_i} \geq 0$
and the minimum value of $\sin{c_i}$ is either $c_2$ or $c_n$. Some of
the terms $\cos{c_i}$ may be negative but their minimum is
$\cos{c_n}$. Now
\begin{flalign*}
A = & p_1 +p_2 \cos{c_2} + \cdots + p_n\cos{c_n} \\
& p_1(1-\cos{c_n})+p_2 (\cos{c_2}-\cos{c_n})+  \cdots + p_{n-1}
(\cos{c_{n-1}}-\cos{c_n}) + \cos{c_n}\\
& \geq p_1(1-\cos{c_n})+ \cos{c_n} = p_1 +(1-p_1)\cos{c_n}
\end{flalign*}
Similarly, we have
\[ B \geq p_n \sin{c_n} \]
We have two cases. First, let the angle $c_n$ lie in the first
quadrant. Then, $\cos{c_n} \geq 0$. Hence,
\[| F(W)|^2 = A^+B^2 \geq  p_1^2 +(1-p_1)^2\cos^2{c_n} +
2p_1(1-p_1)\cos_{c_n} + p_n^2 \sin^2{c_n} \]
}

The next lemma which is very useful for proving important
properties  of $D$ appears as an exercise in \cite{Bhatia}. The proof,
essentially geometric in nature, is not difficult. It is based on the
following fact \cite{Bhatia}.
\begin{emptyT}
Let $A$ and $B$ be normal matrices and let $\norm{A-B}< \epsilon$. If
the disk ${\tt D}(a, \rho)$ with centre $\rho$ and radius $\rho$ in
the complex plane contains $k$ eigenvalues of $A$ then the disk ${\tt
D}(a, \rho+\epsilon)$ contains at least $k$ eigenvalues of $B$.
\end{emptyT}
I do not prove it here as a more general result is given in the
reference quoted above.
\begin{lem} \label{lem:unEigen}
Let $U$ and $V$ be two unitary matrices whose eigenvalues lie on a
semicircle of the unit circle.
Let the eigenvalues $\{a_i\}$( resp.\ $\{b_i\}$)  of $A$( resp.\ $B$)
be labelled counterclockwise. Then,
\[ \max_{i}|a_i-b_i| \leq \norm{A-B} \]
\end{lem}
\commentout{
\begin{proof}
Since the eigenvalues of $U$ and $V$ are assumed to lie on a
semicircle, after multiplying both by a complex number of modulus 1 we
may assume that it is the semicircle on the upper half plane.
The lemma uses the following result( see \cite{Bhatia}).

Using this we see that if we draw small disk of radius $\delta$ around
$a_i$ then the disk \({\tt D} (a_i, \delta+\epsilon), \; \norm{A-B}<
\epsilon\) contains some eigenvalue $b_{k_i}$ of $V$.
\end{proof}
}
Next, I prove several important properties of $D$.
\begin{thm}
For any pair of unitary matrices $U,V$, the function $D(U,V)$
satisfies the following.
\be
\item {Projective invariance} \\
For any complex number $c$ of modulus 1,
\[ D(U,cV) = D(cU,V) = D(U,V) \]
\item {Nonnegative} \\
\(0 \leq D(U,V) \leq 1\) and $D(U,V)= 0$ iff $U=cV,\;
|c|=1$. $D(U,V)=0$ iff $U=cV$ for some complex number $c$ with
$|c|=1$.
\item {Symmetry}
\[D(U,V)=D(V,U)\]
\item {Triangle inequality}\\
For any three unitary matrices $U,V, \text{ and }W$
\[ D(U,V)+D(V,W) \leq D(U,W) \]
\item {Invariance.} $D$ is invariant under left and right translations in the group
  $\cali{U}_n$.
\item
$D(U, V) = 1$ iff there is a unit vector $\alpha$ such that $U\alpha$
  and $V\alpha$ are orthogonal.
\ee
\end{thm}

\begin{proof}
The first three assertions are straightforward consequences of the
definitions. I prove the triangle inequality. Let \(X= \hconj{U}V,
Y=\hconj{V}W \text{ and } Z= \hconj{U}W\). Using the notation in
Theorem \ref{thm:eigenChar} let $F(X)$ (resp.\ \(F(Y), F(Z)\)) denote
the convex polygon spanned by the eigenvalues of $X$ (resp.\
$Y,Z$). Let \(R(X)= \rho^2(0, F(X)) \) and similarly for $F(Y)$ and
$F(Z)$. Since, $Z=XY$ we have
\[ D(U,V)+D(V,W) -D(U,W) = (1-R(X))^{1/2}+(1-R(Y))^{1/2}-
(1-R(XY))^{1/2} \]

\commentout{
To prove triangle inequality it suffices to show that for any three
unitary matrices $X,Y, \text{ and } Z$,
\beq \label{eq:triangle1}
R(X)+R(Y) - R(XY) \leq 1
\eeq
If $R(X), R(Y) \leq 1/2$ then there is nothing to prove. We may
therefore assume that $R(X)\geq 1/2, R(Y)$. Using the projective
invariance property we may further assume that the eigenvalues are
ordered so that 1 is the first eigenvalue of both $X$ and $Y$
corresponding to phase 0. Hence, it follows from Theorem
\ref{thm:eigenChar} that the assumption $R(X)\geq 1/2$ implies that
the eigenvalues of $X$ lie in the first quadrant.
%%Let us first assumethat $R(Y) >0$.
%%It follows agaun from the above theorem that the
%%eigenvalues of $Y$ lie on the upper semicircle.
Now it is clear that the function \(R (X) = \min_{\norm{\alpha}=1}
|\inpr{\alpha }{X}{\alpha}|\) is unitary invariant: \( R(\hconj{U}XU)
= R(X) \) for any unitary matrix $U$. Hence, letting $U$ be the matrix
that diagonalises $X$ we see that it suffices to prove \ref{eq:triangle1}
under the additional assumption that $X$ is diagonal and $Y$
arbitrary. Let From Theorem \ref{thm:eigenChar} \( R(Y) - R(XY)
 }
If both \( D(U,V) = D(1,X) \text{ and } D(V,W) =D(1,Y) \geq 1/2 \)
then there is
nothing to prove. Hence, we assume \(D(U,V) \leq \min( D(V,W), 1/2)
\). This implies that $R(X) \geq \frac{\sqrt{3}}{2}$. Further, we may
also assume that $D(V,W)< 1$. Using the projective
invariance property we may further assume that the eigenvalues are
ordered so that 1 is the first eigenvalue of both $X$ and $Y$
corresponding to phase 0. The preceding assumptions imply that the
eigenvalues $e^{ic_i}$ of $X$ are such that $c_i \leq
\pi/3$. Similarly, if $e^{id_i}$ are the eigenvalues of are the
eigenvalues of $Y$ then $d_i \leq \pi$ as before we order the
eigenvalues counterclockwise such that $c_1=d_1=0$. Then, from Theorem
\ref{thm:eigenChar} it follows that $D(1,X)= \sin{c_n/2}$ and $D(1,Y)=
\sin{d_n/2}$. Let $e^{ih_i}$ be the eigenvalues of $XY$. Using Lemma
\ref{lem:unEigen} we get
\[ |e^{ih_n}-e^{id_n}| = 2\sin{((h_n-d_n)/2)} \leq \norm{Y-XY} =
  \norm{1-X} = 2\sin{(c_n/2)} \]
It follows that \( h_n -d_n \leq c_n \). Now there are two cases.
\be
\item {Case 1}
$h_n \geq \pi$. Then $D(1,XY) =1$. Put $h_n = \pi + c$. Then, $0 \leq \pi
  -d_n \leq c_n -c \leq c_n$. Hence,
\[ D(1,X)+D(1,Y) = \sin{d_n/2}+\sin{c_n/2} \geq
\sin{d_n/2}+\sin{((\pi-d_n)/2)} \geq 1\]
\item {Case 2}
$h_n < \pi$. Then $D(1,XY) =\sin {h_n/2}$ and
\[ D(1,X)+D(1,Y) = \sin{d_n/2}+\sin{c_n/2} \geq \sin{((d_n+c_n)/2)}
  \geq \sin{h_n/2} = D(1,XY) \]
In the last inequality I have assumed that $d_n+c_n < \pi$. Otherwise,
$\sin{d_n/2}+\sin{c_n/2} \geq 1 \geq D(1,XY)$.
\ee
We conclude that the triangle inequality is valid.

If $W$ is any unitary matrix then invariance with respect to
translations means
\[ D(U, V) = D(WU, WV)= D(UW, VW) \]
This follows from
\[\min_{\norm{\psi}=1} \inpr{\psi}{\hconj{U}V}{\psi} =
\min_{\norm{\psi}=1} \inpr{\psi}{\hconj{WU}WV}{\psi} =
\min_{\norm{\psi}=1} \inpr{\psi}{\hconj{UW}VW}{\psi} \]
To prove the last item in the theorem observe that 
\[D(U,V) =1  \text{ iff }  \min_{\norm{\psi}=1} \inpr{\psi}{\hconj{U}V}{\psi} = 0\] 
 That is
if and only if there is some unit vector $\alpha$ such that
$\inpr{\alpha}{\hconj{U}V}{ \alpha}= 0$. But then $U\alpha$ and
$V\alpha$  are orthogonal.
\end{proof}

The triangle inequality for $D$  is proved using the estimates in terms of eigenvalues \ref{thm:eigenChar}. This implies that all the eigenvectors are available so that they may form a basis. But if we restrict to some invariant subspace of the full space a complete set of eigenvectors may not be available. I therefore give an alternative proof of the triangle inequality. 
\begin{lem}
Let \( \cali{U}_n\) act on some finite dimensional Hilbert  space $V$ and let $D$ be defined as 
\[
D(U,V) = \max_{\norm{\psi}} (1-|\inpr{\psi}{U^{\dagger}V}{\psi}|^2 )^{1/2}
\]
Then $D$ satisfies the triangle inequality 
\[D(U,V)\leq D(U,W)+D(W,V) \]
\end{lem}
\begin{proof} 
Let \(R(U)= \min_{\norm{\psi}=1}|\inpr{\psi}{U}{\psi}|^2\). 
From the definitions it follows that we have to show that 
\[ R(\hconj{U}W)+  R(\hconj{W}V)- R(\hconj{U}V) -2 [(1-R(\hconj{U}W))(1-R(\hconj{W}V))]^{1/2} \leq 1 \] 
Since $\hconj{U}V= \hconj{U}W\hconj{W}V$ it suffices to show that for any $U,V\in \cali{U}_n$ 
\beq \label{ineq:RU}
R(U)+  R(V)- R(UV) -2 [(1-R(U))(1-R(V))]^{1/2} \leq 1 
\eeq 
Let $\alpha\in V$ be such that \(R(UV)=  |\inpr{\alpha}{UV}{\alpha}|^2\) and let \(\{\alpha_1=\alpha, \alpha_2, \dotsc, \alpha_n\}\) be an orthonormal basis. Then, 
\[
\begin{split}
R(U)+  R(V)- R(UV)& \leq |\inpr{\alpha}{U}{\alpha}|^2+ |\inpr{\alpha}{V}{\alpha}|^2 - |\inpr{\alpha}{UV}{\alpha}|^2 \\
& = |\inpr{\alpha}{U}{\alpha}|^2+ |\inpr{\alpha}{V}{\alpha}|^2 - |\sum_i \inpr{\alpha_1}{U}{\alpha_i}\inpr{\alpha_i}{V}{\alpha_1}|^2 \\
\end{split}
\]
The last line follows from the resolution of identity $I = \sum_i\pj{\alpha_i}$. Consider the last term. 
\[
\begin{split}
& |\sum_i\inpr{\alpha_1}{U}{\alpha_i} \inpr{\alpha_i}{V}{\alpha_1}|^2  \geq \\
& (|\inpr{\alpha}{U}{\alpha}| |\inpr{\alpha}{V}{\alpha}|- |\sum_{i\neq 1} \inpr{\alpha_1}{U}{\alpha_i}\inpr{\alpha_i}{U}{\alpha_1}|)^2 \geq \\
& (|\inpr{\alpha}{U}{\alpha}| |\inpr{\alpha}{V}{\alpha}|- (\sum_{i\neq 1} |\inpr{\alpha_1}{U}{\alpha_i}|^2)^{1/2} (|\sum_{i\neq 1} |\inpr{\alpha_i}{U}{\alpha_1}|^2)^{1/2})^2 \\
&\geq (|\inpr{\alpha}{U}{\alpha}| |\inpr{\alpha}{V}{\alpha}|- (1-|\inpr{\alpha}{U}{\alpha}|^2)^{1/2}  (1-|\inpr{\alpha}{V}{\alpha}|^2)^{1/2} )^2 \\
& = 1-R(U)-R(V)+2R(U)R(V)-2[R(U)R(V)(1-R(U))(1-R(V))]^{1/2} 
\end{split}
\]
We use Cauchy-Schwartz inequality for getting the fourth line. Using this result we get 
\[
\begin{split}
& R(U)+  R(V)- R(UV) \leq \\
& 2(R(U)+  R(V))-1-2R(U)R(V)+2[R(U)R(V)(1-R(U))(1-R(V))]^{1/2} 
\end{split}
\]
Hence to prove the inequality \ref{ineq:RU} it suffices to show that 
\[
R(U)+R(V)-R(U)R(V) \leq 1+ ((1-R(U))(1-R(V)))^{1/2}(1-(R(U)R(V))^{1/2})
\]
Since $0\leq R(U),R(V) \leq 1$ the above inequality follows from \((1-R(U))(1-R(V))\geq 0\). 
\end{proof}
The special unitary group $S\cali{U}_n$ may be viewed from two
different perspectives. First, as a subgroup of the unitary group
consissting of unitary matrices of order $n$ with determinant 1. The
second point of view is to consider it as factor group. Thus, let
$\cali{D}_n$ consist of all constant unitary matrices. That is
matrices of the form $e^{ic} I_n,\; c \text{ real}$, where $I_N$ is
the unit matrix. Then, $S\cali{U}_n\equiv \cali{U}_n/\cali{D}$. The
algebraic isomorphism is also a topological homeomorphism. We note
that $D(U,V)$ is constant on the cosets of $\cali{D}$. 
Thus we get the following corollary.
\begin{cor}
The function \(D(U,V):S\cali{U}_n \times S\cali{U}_n \rightarrow
\field{R} \) is a metric on $S\cali{U}_n$. 
\end{cor}

Now consider the tensor product $\Complex^m\otimes \Complex^n$ amd the action of 
\(S\cali{U}_n \otimes S\cali{U}_n\) on it. The eigenvalues of an operator of the form 
$U\otimes V$ are given by $u_iv_j$, where $u_i$(resp.\ $v_j$) are eigenvalues of $U$(resp.\ $V$). Then, using theorem \ref{thm:eigenChar} we can show that 
\beq
\begin{split}
D(U_1\otimes U_2 &,V_1\otimes V_2)  = \\
&\min(1,D(U_1,V_1)\sqrt{1-D^2(U_2,V_2)} +D(U_2,V_2)\sqrt{1-D^2(U_1,V_1)})
\end{split}
\eeq
Let us now compare the {\udist} defined above with the standard distance
induced by the {\bf sup}-norm. Let $U,V\in \cali{U}_n$ . Then,
\[ \norm{U-V} = \norm{U(1-\hconj{U}V)} = \norm{I-\hconj{U}V} \]
Write $W = \hconj{U}V$. Since $I-W$ is normal \(\norm{1-W}=
\max\{|\lambda| | \lambda \text{ an eigenvalue of } 1-W\}\). If
$\{e^{ic_k} | k=1,\dotsc, n\} $ are the eigenvalues of ordered
counterclockwise so that $0 \leq c_1 \leq c_2 \cdots \leq c_n <
2\pi$. Let $c_k$ be the phase angle that is closest to $\pi$. Then,
\beq \label{eq:supdistU}
\norm{U-V}= 2\sin{(c_k/2)}
\eeq
Note that, for real $x$,
$\norm{I-e^{ix}I}= 2|\sin{x/2}|$. Thus, if $ |\sin{x/2}|$ is
relatively large then $e^{ix}I$ can not be close to $I$. But the
operator $e^{ix}I$ is simply multiplies all the states by a constant
phase and hence leaves the projective space of quantum states
invariant. We see that the distance induced my the {\bf sup}-norm does
not have projective invariance. For example, the operators $I$ and
$-I$ have maximal distance( =2) between them. The same situations
exists for the distance induced by the Frobenius or trace norm on
matrices defined by $|A|_F = \tr(\hconj{A}A)$. The {\udist} however has
manifest projective invariance. This is one of the reasons for
the introduction of the metric $D(U,V)$. However, the latter is
closely related to the standard distance.
\begin{propn}
Let $U$ and $V$ be unitary opertors. Then
\[ D(U,V) = \frac{1}{2} \norm{U-e^{ix}V} \]
for some real $x$.
\end{propn}

\begin{proof}
We have \(\norm{U-V}= \norm{1 -W} \text{ where } W=
\hconj{U}V\). First, assume that the eigenvalues of $W$ lie on a
semicircle. By multiplying $W$ with appropriate factor $e^{ix}$ we may
assume that
it is the upper semicircle and that 1 is an eigenvalue of $W$. If we
order the eigenvalues $e^{ic_1}=1, \dotsc, e^{ic_n}$ counterclockwise
so that \( 0=c_1 \leq c_2\leq \dotsc \leq \pi\), we have $D(U,V)=
\sin{(c_n/2)}$( see Theorem \ref{thm:eigenChar}) and $\norm{U-e^{ix}V}=
2\sin{(c_n/2)}$( equation \ref{eq:supdistU}). The lemma follows.

In the second case if the eigenvalues of $W$ span an arc which {\em
  includes} a semicircle then $D(U,V)$ attains its maximal value
  1 and we multiply $V$( and $W$) with a factor  $e^{ix}$ such that -1
  is an eigenvalue of $W$. Then \( \norm{U-e^{ix}V} =2\) and the proof is
  complete.
\end{proof} 
\section{Examples and Applications} 
The metric $D$ has some obvious physical interpretations. If we think of $U,V$ as evolution operators. Thus, we write \(U(t,t_0)\text{ and } V(t,t_0) \) to indicate the time dependance. If $\ket{\psi_0}$ is the initial state vector then 
let \(\ket{\psi_t}= U(t,t_0)\ket{\psi(t_0)} \text{ and }  \ket{\psi'_t}= V(t,t_0)\ket{\psi(t_0)} \) be the two vectors at time $t$ corresponding to the two evolution operators. Then the square of {\udist} between $U$ and $V$ at time $t$ is maximum of the quantity ($1-$ the transition probability between $ \ket{\psi_t}$ and $\ket{\psi'_t}$), the maximum being taken over 
all initial states. In the case when the Hamiltonian is independent of time we could also visualize this as two consecutive operations on the same system. First, the evolution operator $U(t_0,t)$ followed by $V{\dagger}(t,2t-t_0)$ 

A quantum circuit is a unitary operator composed of unitary operators of order less than or equal to some fixed number $k$. If the operators $U_c$ and $V_c$ represent two such circuits then we say that $V_c$ $\epsilon-$approximates $U_c$ if $D(U_c,V_c) \leq \epsilon$. We may 
verify such a claim as follows: 
\be
\item
Apply inputs in arbitrary state $\ket{\alpha}$ to $V_c$. 
\item 
Apply the output of $V_c$ to the {\em output} gates of $U_c$.  
\em
Do a projective measurement with respect to the pair of projection operators $P_{\alpha}\equiv\pj{\alpha}$ and $P_{\alpha}^{\perp}$ at the {\em input} of 
$U_c$. We are actually applying $U^{\dagger}=U^{-1}$. 
\item 
If the estimated probability for the oucome $\ket{\alpha}$ is $\geq 1-\epsilon$ for all states 
then $V_c$ $\epsilon-$approximates $U_c$. 
\ee
There are some unsatisfactory aspects to the simplistic approach outlined above. First, the estimated probability based on observed relative frequencies is not the actual probability. This can be rectified by giving upper bounds to the diffrence between the two. The requirement 
that the probabilities be calculated for all states is impossible to satisfy. It is however sufficient to verify that 
$D_{\psi}(U_c,V_c)$ is sufficiently small for all vectors $\psi$ in $n+1$ independent bases, where $n$ is the dimension of the underlying Hilbert space $H$. Here, ``independent bases'' means the following. Let, \(\cali{B}^k= \{\alpha^k_1, \dotsc, \alpha^k_n\}, \; k=1, \dotsc, n+1\) be orthonormal bases in $H$ and let $P^k_i = \pj{\alpha^k_i} $ be the corresponding projection operators onto $\alpha^k_i$. The hermitian operators $P^k_i$ have trace 1. Now a general quantum state is density operator, that is, a positive definite operator $\rho$ of trace 1. If $I$ denotes the identity operator then $\rho - I/n$ is a hermitian operator with trace 0. We say that the bases $\cali{B}^k$ are independent if the traceless operators $P^k_i -I/n,\; i=1,\dotsc, n-1$ are linearly independent. Then, the latter span the space of traceless hermitian operators. Hence, $\rho-I/n$ can be written as a linear combination of the operators $P^k_i -I/n,\; i=1,\dotsc, n-1$. The reader may see \cite{Ivanovic} or \cite{Patra05} for details. Then, the state $\rho$ is uniquely determined by the transition probabilities $\tr(\rho P^k_i)$. Similarly, the state $W\cdot \rho$ is determined by the probabilities $\tr(W\cdot \rho P^k_i)$. Hence, if $\tr(\rho P^k_i)$ are close to $\tr(W\cdot \rho P^k_i)$ then $\rho$ and $U\cdot \rho$ will be close. In particular, if $\rho=\pj{\psi}$ is a pure state then, $|\inpr{\psi}{W}{\psi}|^2$ is close to 1 implying that $D(U_c,V_c)$ is ``small''. We can formalize the abive arguments in case of specific bases( e.\ g.\  {\em mutually unbiased bases}) and get an upper bound on $D(U_c, V_c)$. 

\subsection{ Quantum Search Algorithms} 
In this subsection I discuss application of the {\udist} to a class of algorithms known as quantum search algorithms. The name derives from the fact that these algorithms can be adapted to the problem of search in an unordered database. I give below a generic description of the algorithm. Let $H= \Complex^N$ be a Hilbert space of dimension $N$. We are given a ``standard'' $\cali{B}$ basis in $H$. Write the elements of $\cali{B}$ as \(\ket{1}, \dotsc, \ket{N}\}\).  Suppose that we are given a ``blackbox'' or oracle unitary transformations $O_a= I- 2\pj{a},\; a=1, \dotsc, N$. We use  a sequence of unitary  operators \(U_k, \ldots, U_1\) interleaved with queries to the oracle. Thus, the quantum circuit is given by the unitary operator 
\beq
F_{a,k} \equiv U_k O_a U_{k-1}O_a \cdots U_1O_a 
\eeq
such that the probability of obtaining the result $\ket{a}$ in a measurement in the basis $\cali{B}$ is greater than 1/2. That is, given the initial state $\psi$
\beq \label{eq:spec}
 |\inpr{a}{F_{a,k}}{\psi}|^2 > \frac{1}{2} +c ,\; c> 0
\eeq
where $c$ is positive constant that is independent of $N$. 
We assume that the probability distribution over the integers \(J_N = \{1, \dotsc, N\}\) is uniform. This implies that the probability of the blackbox operator being $O_a$ is equal($=1/N$) for all $a\in J_N$. Then we may suppose that $\psi$ is the totally symmetric state vector. 
\[ \psi= \frac{1}{\sqrt{N}} \sum_i \ket{i} \] 
Let $G_a$ be the unitary operator that acts on the ``plane'' \( {\tt T}\equiv \text{ Span } \{\psi,\ket{a}\}\) leaving all vectors perpendicular to ${\tt T}$ and permutes $\psi$ and $\ket{a}$. Then the probability specification \ref{eq:spec} can be written as 
\[ D_{\psi}(G_a, F_{a,k}) > \frac{1}{2} \]  
The integer $k$, which gives the query complexity is also an estimate of the circuit size which is related to time complexity. Let us calculate bounds for $k$. Our method of getting these estimates differs from the original one given in \cite{Vaz} and illustrates the use of the concepts introduced ealier. 

Let $V_k = U_k\cdots U_1$. Let \(\Phi = \psi\otimes \psi \in H\otimes H\). Define the following operators on $H\otimes H$ by their action on the basis \( \{ \ket{a}\otimes \ket{b} \text{  in } H\otimes H \}\). 
\begin{gather}
F'_{k}( \ket{a}\otimes \ket{b})= \ket{a}\otimes F_{a,k}\ket{b}\quad  V'_k = I\otimes V'_k \\ \quad P' (\ket{a}\otimes\ket{b}) = e^{i\lambda_a}\ket{a} \otimes G_a \ket{b} 
\end{gather}
In the above formula $\lambda_a $ are real numbers to be specified. We may visualize the operators $F'_{k} \text{ and } P'$ as controlled operation such that if the first ``qunit'' is $a$ then the $F_{a,k}$ and $G_a$ are respectively applied to the second. It is easy to verify that all the operators are unitary. Now, it follows from the basic relation  \ref{eq:basic_relation} that, for some real $x$, 
\beq
\frac{1}{2} \norm{F'_{k}\Phi- e^{ix}V'_k\Phi}^2 \leq D^2_{\Phi}(F'_{k}, V'_k) \leq \norm{F'_{k}\Phi- V'_k\Phi}^2
\eeq
We use these relations to get lower and upper estimates of $D_{\Phi}(F'_{k}, V'_k)$. First we note that 
\[
\begin{split}
\norm{F'_{k}\Phi- V'_k\Phi}^2 & = \norm{\frac{1}{\sqrt{N}} (F'_{k}-V'_k)\sum_a \ket{a}\otimes \psi} \\
                              &= \frac{1}{N}\norm{ \sum_a (F_{a,k}-V_k)  \psi} ^2
\end{split}
\]
Using a straightforward calculation( see \cite{NC}) we get \(\norm{ \sum_a (F_{a,k}-V_k)  \psi} ^2 \leq 4k^2\). Hence,
\beq \label{eq:GroverUp}
D^2_{\Phi}(F'_{k}, V'_k) \leq \frac{4k^2}{N}
\eeq
Next, since \( D_{\Phi}(F'_{k}, V'_k) \geq 1/\sqrt{2}\norm{F'_{k}\Phi- e^{ix}V'_k\Phi} \geq 1/\sqrt{2}(\norm{(zP'-e^{ix} V'_{k})\Phi}- \norm{(zP'- F'_k)\Phi} )\), where $z$ is a complex number of modulus 1. By choosing an appropriate $z$ and using the projective invariance of the function $D_{\Phi}$, we get \( \norm{(zP'-e^{ix} V'_{k})\Phi} \geq D_{\Phi}(P',V'_{k}) \text{ and } 1/\sqrt{2} \norm{(zP'- F'_k)\Phi} \leq D_{\Phi} (P',F'_k) \). 
Hence, 
\[ D_{\Phi} (F'_k, V'_k) \geq \frac{1} {\sqrt{2}}D_{\Phi}( P',V'_{k})- D_{\Phi} (P',F'_k) \] 
Using the definition of $D_{\Phi}$ and the operators, we get 
\[   
\begin{split}
D_{\Phi}( P', V'_{k}) & = (1-  | \frac{1}{N}\sum_a e^{-i\lambda_a} \inp{a}{V_k\psi}|^2 )^{1/2} \\ 
                     & \geq (1- \frac{\sum_a |\inp{a}{V_k \psi}|^2}{N})^{1/2}= (1-\frac{1}{N})^{1/2} 
\end{split} 
\]
In the second step we use Cauchy-Schwartz inequality and the fact that \( \sum_a |\inp{a}{V_k\psi}|^2=1\). On the other hand we also have 
\[ 
D_{\Phi} (P',F'_k)  = (1 - | \sum_a e^{-i\lambda_a} \frac{\inp{a}{F_{k,a}\psi}}{N}|^2)^{1/2} \]
We now define \(e^{i\lambda_a}= \inp{a}{V_k}{\psi}/|\inp{a}{V_k\psi}|\) if $|\inp{a}{V_k\psi}|\neq 0$ and 1 otherwise. Then,  from equation \ref{eq:spec} it follows that \(e^{-i\lambda_a}\inp{a}{V_k\psi}= |\inp{a}{V_k \psi}|\geq (1/2+c)^{1/2}\). Hence, 
\[
D_{\Phi}( P', F'_{k}) \leq (\frac{1}{2}-c)^{1/2} 
\]
Combining the estimates for \(D_{\Phi}( P', V'_{k}) \text{ and } D_{\Phi}( P', F'_{k})\) we get 
\[ D_{\Phi} (F'_k, V'_k) \geq \frac{1}{\sqrt{2}} (1-\frac{1}{N})^{1/2} - (\frac{1}{2}-c)^{1/2} \geq \frac{1}{\sqrt{2}} ((1-\frac{1}{N})^{1/2} - 1+c) \]
As we are only interested in asymptotic behaviour, by taking $N$ large enough we have \( (1-\frac{1}{N})^{1/2} - 1+c \geq c/\sqrt{2} \). Hence, 
\beq
D_{\Phi} ^2 (F'_k, V'_k) \geq c^2 /4 .
\eeq 
Combining the the two bounds for $D_{\Phi} (F'_k, V'_k) $ we get $k^2/N \geq c^2/4 $. That is, $k =O( \sqrt{N})$. The complexity of the Grover quantum search algorithm is $O(\sqrt{N})$ and it is the best possible. 
\section{Some Estimates and Generalizations}
In this section I give some estimates of the metric $D$ in special cases. Let us estimate the  {\udist} for some special unitary operators. The CNOT-gate $C$ \cite{NC} is a unitary opertor on 4-dimensional Hilbert space $C$, such that 
\begin{gather*} 
C\ket{0}\tensor\ket{0}= \ket{0}\tensor\ket{0} \quad C \ket{0}\tensor\ket{1} \ket{0}\tensor\ket{1} \\
C\ket{1}\tensor\ket{0}= \ket{1}\tensor\ket{1} \quad C \ket{1}\tensor\ket{1} \ket{1}\tensor\ket{0} 
\end{gather*}

The states $\ket{0},\ket{1}$ are any pair of 2-dimensional orthogonal vectors. It can be  shown that 
\beq
D(C, U\otimes V) \geq 1/2 
\eeq
where $U$ and $V$ are arbitrary unitary operators in 2-dimensions. In fact we can show more. 
Namely, that the operator $C$ is at ditsance $\geq 1/2$ from the subgroup of $H_4$ of $SU_4$ generated by the permutation( swap) operators, and the product matrices of the form $U\otimes V$ and this distance is maximal. That is, the $CNOT$ gate is optimal for entanglement of two qubits, a well-known result. We may therefore define a measure on the unitary operators $SU_N, N=2^n$ as follows. Let $H_N$ be the subgroup that leaves the set of product states invariant. Then for any unitary operator $U$,
\beq
\rho_E(U,H_N)\equiv D(U,H_N)=\inf_{V\in H_N} D(U,V) 
\eeq
It is conjectured that $H_N$ is generated by single qubit operators and permutations. 

We have defined the distance $D$ on the group of unitary operators via their natural representation, i.e.\ $SU_n$ on $\Complex^n$, but we could extend it to any action of the group. For example, consider the action of $SU_n$ the set of density operators: $U\cdot \gamma = U\gamma \hconj{U}$. But then, we have to be careful in checking the triangle inequality. Moreover, if we try a naive extension to the infinite-dimensional case we have to deal with convergence 
issues. I aim to deal with these issues in future. 
\bibliographystyle{phaip}

\end{document}